\documentclass[12pt]{article}
\usepackage{amsmath,amssymb,array}
\usepackage{graphicx}

\numberwithin{equation}{section}
\def\p{\partial}

\begin{document}

\begin{titlepage}
\renewcommand{\thefootnote}{\fnsymbol{footnote}}

\begin{center}

\begin{flushright}arXiv: 1003.0558\end{flushright}
\vspace{3.5cm}

\textbf{\Large{Linear Confinement for Mesons and Nucleons\\[0.5cm] in AdS/QCD}}\vspace{2cm}

\textbf{Peng Zhang} \\[0.5cm]

\textsf{E-mail: pzhang@bjut.edu.cn}\\[0.5cm]

\emph{Institute of Theoretical Physics, College of Applied Sciences, \\
      Beijing University of Technology, Beijing 100124, P.R.China}

\end{center}\vspace{1.5cm}

\centerline{\textbf{Abstract}}\vspace{0.5cm}
By using a new parametrization of the dilaton field and including a cubic term in the bulk scalar potential,
we realize linear confinement in both meson and nucleon sectors within the framework of soft-wall AdS/QCD.
At the same time this model also correctly incorporate chiral symmetry breaking. We compare our resulting
mass spectra with experimental data and find good agreement between them.

\end{titlepage}
\setcounter{footnote}{0}

\section{Introduction}

Quantum chromodynamics has been identified as the correct theory of
strong interactions for nearly forty years. It describes the process
with large transfer momenta, like deeply inelastic scattering
experiment,  quite well. Due to the property of asymptotic freedom,
the coupling constant is small in these cases, so the perturbation
theory is reliable. As the characteristic energy lowers, the theory
becomes strongly coupled and difficult to directly describe the
behavior of mesons and baryons. People then develop various
effective models exhibiting some important phenomena, such as linear
confinement and chiral symmetry breaking, which are believed to be
the features of low energy QCD. An excellent reference in this field
is \cite{Nar}.

Recently a new methodology, called ``AdS/QCD'' or more generally
``holographic QCD'', appears \cite{EKSS, DP} and has been
intensively studied. It is motivated by the AdS/CFT correspondence
in string theory \cite{M, GKP, W1}, which identifies the
$\mathcal{N}=4$ four dimensional supersymmetric gauge theory on one
side with the type IIB string theory in $AdS_5 \times S^5$ with
Ramond-Ramond 5-form fluxes on the other side. Based on some basic
setups in \cite{W2, KMMW}, the authors of \cite{SS} construct a
semi-realistic holographic dual of QCD from string theory. By use of
the general dictionary of the correspondence, AdS/QCD, the bottom-up
counterpart of the above stringy methods, directly introduce fields
corresponding to important operators in QCD in the bulk. This
method, compared to those top-down constructions, has much more
flexibilities to implement the phenomenological studies.

Chiral symmetry breaking can be realized in the so-called hard-wall
model where we only use a slice of $AdS_5$ with an IR cutoff and the
QCD scale is related to the cutoff. Low-lying meson spectra and
couplings fit well with the experimental data. Baryons can also be
incorporated in this model, by introducing a pair of 5d Dirac
spinors for spin-$\frac{1}{2}$ nucleons \cite{HIY} \footnote {There
is also a 5d skyrmion-like baryon model, see \cite{PW1, PW2}.}, and
Rarita-Schwinger fields for spin-$\frac{3}{2}$ $\Delta$ resonances
\cite{AHPS}. However the main problem in the hard-wall model is the
deviation of excited state masses compared with data, which should
grow approximately linearly due to the confinement of quarks. To
realize linear confinement, \cite{KKSS} introduce a soft-wall model
where the IR cutoff is removed. A quadratic background dilaton field
is used to suppress the Kaluza-Klein mode functions at infinity, and
gives a linear growth spectra for the vector mesons. However this
model entangles the explicit and spontaneous breaking of chiral
symmetries. This problem is resolved in \cite{GKK} by using a
modified dilaton and a quartic term in the bulk scalar potential.
There have been lots of works on AdS/QCD. For instance, studying of
meson properties \cite{DaRold:2005vr}-\cite{Zuo}; relation with
light-front dynamics \cite{BdT}; nucleons in the hard-wall model
\cite{MT}-\cite{Z}, and many others.

The aim of this paper is to realize linear confinement in both meson
and nucleon sectors, and correctly incorporate chiral symmetry
breaking\footnote{There are already some works about this subject in
different models, see e.g. \cite{Forkel:2007cm, BdT(N)}.}. We will
use a new parametrization of the background dilaton field, which is
still quadratic growth at infinity to generate a linear growth meson
spectra. While in the nucleon sector the dilaton cannot supply a
soft-wall to discretize the spectrum, since the bulk action for
Dirac spinors is first order.  However we can obtain a linear growth
nucleon spectra if the v.e.v. of the bulk scalar, which is the
holographic dual of the chiral condensate, is asymptotically
quadratic. This can be obtained by a cubic term in the bulk scalar
potential. The remaining parts of this paper are organized as
follows: In section 2 we will introduce the modified soft-wall model
for meson sectors and show that how the new dilaton and the cubic
term can lead to a linear growth meson spectra. In section 3 we
describe how to realize linear confinement in the nucleon sector by
using a new parametrization of the v.e.v of the bulk scalar. In
section 4 we summarize our results and discuss some further issues.

\section{Meson sector}

We will consider a modified version of the soft-wall AdS/QCD model first introduced in \cite{KKSS}. The background geometry is
chosen to be 5d anti-de Sitter space
\begin{eqnarray}
ds^2=\,G_{MN}\,dx^M dx^N=a^2(z)\,(\,\eta_{\mu\nu}dx^{\mu}dx^{\nu}-dz^2)\,,\quad  0 \leq z < \infty\,,
\end{eqnarray}
where $a(z)=1/z$. Our convention of the Minkowski metric $\eta_{\mu\nu}$ is diag(1,-1,-1,-1).
There is also a background  dilaton field $\Phi(z)$.
The bulk action for the meson sector is
\begin{eqnarray}
S_M=\int d^5x \sqrt{G}\,e^{-\Phi} \left\{-\frac{1}{4g_5^2}(\,\|F_L\|^2+\|F_R\|^2)+\|DX\|^2-m_X^2\|X\|^2-\lambda\,\|X\|^3\right\}\,. \label{SM}
\end{eqnarray}
The norm of a matrix $A$ is defined as $\|A\|^2=\mathrm{Tr}(A^\dag A)$.
Here we include a cubic term in the bulk scalar potential. This term allows the dilaton $\Phi(z)$ and $X_0(z)$, the v.e.v. of $X$,
to be both quadratic growth as $z\rightarrow\infty$. These are crucial for obtaining linear spectra for meson and nucleon sectors.
We will discuss this point more carefully below.

Other terms in (\ref{SM}) are as usual: $F_L$ and $F_R$ are the field strength of
two $SU(2)$ gauge fields, $L$ and $R$, which are holographic duals of the left-handed and right-handed chiral currents.
The coupling constant $g_5^2=12\pi^2/N_c=4\pi^2$ is fixed by matching the two-point function of vector currents.
The bulk scalar $X$, with $m_X^2=-3$, is dual to the chiral condensate, which transforms in the bi-fundamental representation of
$SU(2)_L \times SU(2)_R$. The covariant derivative is defined as $D_M{X}=\p_M{X}-iL_M{X}+i{X}R_M$. The trace
is taken in the fundamental representation of $SU(2)$.

\subsection{Bulk scalar VEV and dilaton background}

The bulk scalar $X$ is expected to obtain a $z$-dependent v.e.v. as follows
\begin{eqnarray}
\langle{X}\rangle=X_0(z) \begin{pmatrix} 1 & 0 \\ 0 & 1 \end{pmatrix}.
\end{eqnarray}
From the bulk action (\ref{SM}) it can be seen that the possible $X_0(z)$ and the dilaton $\Phi(z)$ are related with
each other through the EOM
\begin{eqnarray}
\p_z(\,a^3 e^{-\Phi}\p_z X_0)-a^5 e^{-\Phi}(m_X^2X_0+\frac{3}{2}\lambda X_0^2\,)=0\,. \label{EOMX}
\end{eqnarray}
The dilaton appears here because the action is second order and we need to do a integration by parts.
This is a nonlinear differential equation for $X_0$, and difficult to solve for given dilaton profile.
Here we follow the strategy of \cite{GKK}, i.e. first choose a parametrization of $X_0$ satisfying proper
boundary conditions, then using (\ref{EOMX}) to determine to background dilaton $\Phi$ up to a integral constant
\begin{eqnarray}
\Phi'(z)=\,\frac{1}{a^3 X_0'}\,((\,a^3X_0\,')\,'-a^5(m_X^2X_0+\frac{3}{2}\lambda X_0^2\,))\,. \label{phi'}
\end{eqnarray}
Here the prime denotes the derivative with respect to $z$.
We assume the asymptotic behavior of $X_0$ to be
\begin{eqnarray}
&&X_0\, \sim\, \frac{1}{2}\,(m_q\zeta z+\frac{\sigma}{\zeta}z^3)\,,\quad  z\rightarrow0\,;\\[0.2cm]
&&X_0\, \sim\, \frac{1}{2}\,\gamma z^2\,,\quad\hspace{1.7cm}  z\rightarrow\infty\,. \label{Xinf}
\end{eqnarray}
The normalization $\zeta=\sqrt{3}/(2\pi)$, see \cite{CCW}. The parameter $m_q$ and $\sigma$ are
interpreted as the quark mass and chiral condensate. The second equation is needed for having a linear nucleon spectra,
which becomes clear when we discuss the nucleon sector in the next section. To satisfy these two boundary conditions
we choose the parametrization of $X_0$ as follows
\begin{eqnarray}
X_0(z)=\frac{\,Az+Bz^3}{\sqrt{1+C^2z^2}\,}\,\,.
\end{eqnarray}
The relations between $m_q,\sigma,\gamma$ and $A,B,C$ are
\begin{eqnarray}
m_q=\frac{2A}{\zeta}\,,\qquad \sigma=2\zeta\,(B-\frac{1}{2}AC^2)\,,\qquad \gamma=\frac{2B}{C}\,.\label{rel}
\end{eqnarray}
By use of the relation (\ref{phi'}) we can see that the asymptotic behavior of the dilaton field $\Phi(z)$ is indeed $O(z^2)$ as $z\rightarrow\infty$.
This will give us an asymptotically linear meson spectra.

We will determine the parameters $A,B,C$ and $\lambda$ by fitting the experimental data of the vector and axial-vector meson masses
under the constraint from the Gell-Mann-Oakes-Renner relation $2m_q\sigma=f_{\pi}^2m_{\pi}^2$ with $m_{\pi}=139.6$ Mev and $f_{\pi}=92.4$ Mev.
The fitting results are: $A=0.82$ Mev, $B=(298.9\,\mathrm{Mev})^3$, $C=(1353.62\,\mathrm{Mev})^2$, and
\begin{eqnarray}
\lambda=41.48.
\end{eqnarray}
Then by using (\ref{rel}) we can obtain the values of $m_q,\sigma,\gamma$ as
\begin{eqnarray}
m_q=5.95\,\mathrm{Mev}\,,\quad \sigma=(241\,\mathrm{Mev})^3\,,\quad
\gamma=29.1\,\mathrm{Mev}.
\end{eqnarray}
In the next two subsections we will use them to calculate the mass spectra of vector and axial-vector mesons.

\subsection{Vector mesons}
From two bulk $SU(2)$ gauge fields $L_M$ and $R_M$, we can introduce $V_M=(L_M+R_M)/\sqrt{2}$.
The vector $\rho$ meson and its radial excitations are identified, in this model, as the Kaluza-Klein
modes of $V_\mu$. Decompose the field $V_\mu(x,z)$ as
\begin{eqnarray}
V_\mu(x,z)=\sum_n\, \rho_\mu^{(n)}(x)\,f_V^{(n)}(z)\,.
\end{eqnarray}
Insert this expression into the action (\ref{SM}), with the gauge condition $V_z=0$, we can get the
equation of motion for $f_V^{(n)}$ as
\begin{eqnarray}
-\p_z^2f_V^{(n)}+\omega'\p_zf_V^{(n)}=m_V^{(n)\,2}f_V^{(n)}\,,
\end{eqnarray}
where $\omega=\Phi(z)-a'/a=\Phi(z)+\log{z}$. After a Liouville transformation $f_V^{(n)}=e^{\omega/2}\psi_V^{(n)}$,
we can write the above equation in a Schr\"{o}dinger form
\begin{eqnarray}
-\p_z^2 \psi_V^{(n)} + \left(\frac{1}{4}\omega'^2-\frac{1}{2}\omega''\right)\psi_V^{(n)}=m_V^{(n)\,2}\psi_V^{(n)}\,.
\end{eqnarray}
The effective potential is $O(z^{-2})$ as $z\rightarrow0$ and $O(z^2)$ as $z\rightarrow\infty$.
Therefore the resulting spectrum is asymptotically linear. We can numerically solve this eigenvalue problem.
The UV boundary condition is $\psi_V^{(n)}(z\rightarrow0)=0$. The boundary condition in deep IR (i.e. $z\rightarrow\infty$)
is fixed by the requirement of normalizability: $\int_\Lambda^\infty e^{-\Phi}|\,f_V|^2dz<\infty$, with $\Lambda$ being some
large number. Due to the relation between $f_V$ and $\psi_V$, this is equivalent to $\int_\Lambda^\infty |\,\psi_V|^2dz<\infty$.
Since the effective potential is $O(z^2)$ at infinity, the wave function $\psi_V \sim e^{-\kappa z^2}$ as $z\rightarrow\infty$
for some $\kappa>0$. This automatically implies $\p_z\psi_V^{(n)}(z\rightarrow\infty)=0$ used in e.g. \cite{KKSS, GKK}.

In Table \ref{rho} we show the predicted values of our model and corresponding experimental data \cite{PDG}.
We can see that except the ground state the agreement with the data for other excited states are
all within 10\%. The average error is 6.4\%.
\begin{table}
\centering
\begin{tabular}{|c|c|c|c|c|c|c|}
\hline
$\rho$                &  0    &   1  &  2   &  3   &  4   &  5   \\
\hline
$m_{\mathrm{exp}}$(Mev) & 775.5 & 1465 & 1720 & 1909 & 2149 & 2265 \\
\hline
$m_{\mathrm{th}}$ (Mev)  & 952.6 & 1349 & 1653 & 1909 & 2134 & 2338 \\
\hline
error              &22.8\% &7.9\% &3.9\% &0.0\%   &0.7\% &3.2\% \\
\hline
\end{tabular}
\caption{\small{The experimental and theoretical values for vector meson masses.
The average error is 6.4\%.}}\label{rho}
\end{table}

\subsection{Axial-vector mesons}
We define another gauge field $A_M=(L_M-R_M)/\sqrt{2}$. The axial-vector mesons are identified as the Kaluza-Klein
modes of $A_\mu$. Decompose the field $A_\mu(x,z)$ as
\begin{eqnarray}
A_\mu(x,z)=\sum_n\, a_\mu^{(n)}(x)\,f_A^{(n)}(z)\,.
\end{eqnarray}
Similarly choose the gauge condition $A_z=0$, the equation for $f_A^{(n)}(z)$ is
\begin{eqnarray}
-\p_z^2f_A^{(n)}+\omega'\p_zf_A^{(n)}+4\,g_5^2\,a^2X_0^2f_A^{(n)}=m_A^{(n)\,2}f_A^{(n)}\,,
\end{eqnarray}
Note that, unlike the vector mesons, this axial-vector field couples to the scalar v.e.v. $X_0(z)$,
which produce a $z$-dependent mass term in the equation of motion. By using the same substitution
$f_A^{(n)}=e^{\omega/2}\psi_A^{(n)}$ as before, the Schr\"{o}dinger equation for axial-vector mesons is
\begin{eqnarray}
-\p_z^2 \psi_A^{(n)} + \left(\frac{1}{4}\omega'^2-\frac{1}{2}\omega''+4\,g_5^2\,a^2X_0^2\right)\psi_A^{(n)}=m_A^{(n)\,2}\psi_A^{(n)}\,.
\end{eqnarray}
The effective potential for the axial-vector mesons is $O(z^{-2})$ as $z\rightarrow0$ and $O(z^2)$ as $z\rightarrow\infty$.

Since the term $4g_5^2a^2X_0^2$ is also quadratic growth at infinity in our model, the asymptotic slope of the axial-vector
meson spectra is different from that of the vector sector. This fact is a direct consequence of the asymptotic behavior
of the dilaton $\Phi$ and the vev $X_0$, which are necessary to obtain the linear spectra in both meson and nucleon sectors.
However it has been argued \cite{SV, GKK}, based on data, that these slopes should become asymptotically equal. It seems to the author that,
if we realize nucleon states through the method of \cite{HIY}, there would be a clash between an asymptotically linear nucleon
spectra and asymptotically equal slopes for mesons.

We can numerically solve this eigenvalue problem under the similar boundary condition $\psi_A^{(n)}(z\rightarrow0)=0$
and $\p_z\psi_A^{(n)}(z\rightarrow\infty)=0$. Since the effective potential for axial-vector mesons is also $O(z^2)$ at infinity,
the discussion about the IR boundary condition given in the above subsection also applies here. In Table \ref{a1} we show the predicted
values of our model and corresponding experimental data \cite{PDG}. We can see that except the ground state the agreement with the data
for other excited states are all within 10\%. The average error is 6.7\%.
\begin{table}
\centering
\begin{tabular}{|c|c|c|c|c|c|c|}
\hline
$a_1$                &  0    &   1  &  2   &  3   &  4      \\
\hline
$m_{\mathrm{exp}}$(Mev) & 1230 & 1647 & 1930 & 2096 & 2270  \\
\hline
$m_{\mathrm{th}}$ (Mev)  & 1065 & 1520 & 1867 & 2158 & 2414  \\
\hline
error              &13.4\% &7.7\% &3.3\% &2.9\%   &6.3\%  \\
\hline
\end{tabular}
\caption{\small{The experimental and theoretical values for axial-vector meson masses.
The average error is 6.7\%.}}\label{a1}
\end{table}

\section{Nucleon sector}

Now we will turn to the nucleon sector. As in \cite{HIY} we realize spin-$\frac{1}{2}$ nucleons
by introducing a pair of 5d Dirac spinors $\Psi_{1,2}$ which couple with the corresponding baryon operators
on the UV boundary $z=0$. The bulk action for the nucleon sector is
\begin{eqnarray}
S_N&=&\int d^5x \sqrt{G}\,e^{-\Phi}\,\mathrm{Tr}(\,\mathcal{L}_K+\mathcal{L}_I) \,, \nonumber\\
\mathcal{L}_K&=&i\bar{\Psi}_1\Gamma^M\nabla_M\Psi_1+i\bar{\Psi}_2\Gamma^M\nabla_M\Psi_2
                -m_{\Psi}\bar{\Psi}_1\Psi_1+m_{\Psi}\bar{\Psi}_2\Psi_2 \,, \nonumber\\[0.2cm]
\mathcal{L}\,_I&=&-g\,\bar{\Psi}_1X\Psi_2-g\,\bar{\Psi}_2X^\dag\Psi_1\,.
\end{eqnarray}
Here $\Gamma^M=e^M_A\Gamma^A=z\delta^M_A\Gamma^A$, and $\{\Gamma^A,\Gamma^B\}=2\,\eta^{AB}$
with $A=(a,5)$. We can choose the representation as $\Gamma^A=(\gamma^a, -i\gamma^5)$.
The covariant derivatives for spinors are
\begin{eqnarray}
\nabla_M \Psi_1&=&\p_M\Psi_1+\frac{1}{2}\,\omega^{AB}_M\Sigma_{AB}\Psi_1-iL_M\Psi_1 \,,\\
\nabla_M \Psi_2&=&\p_M\Psi_2+\frac{1}{2}\,\omega^{AB}_M\Sigma_{AB}\Psi_2-iR_M\Psi_2 \,.
\end{eqnarray}
Here $\Sigma_{AB}=\frac{1}{4}[\Gamma_A,\Gamma_B]$, and the nonzero components of the spin
connection $\omega^{AB}_M$ is $\omega^{a5}_\mu=-\omega^{5a}_\mu=\frac{1}{z}\,\delta^a_\mu$.
The mass $m_{\Psi}$ can be related to the dimension of the baryon operator. Its classical value is 5/2.
Since the operator may has nonzero anomalous dimension, we can also treat $m_{\Psi}$ as a free parameter as suggested in \cite{HIY}.
This will improve our fitting result.
The Yukawa coupling between $\Psi$'s and $X$ is required by
the chiral symmetry breaking in the nucleon sector. This term will also give us a parity doublet
pattern in the spectrum. In the nucleon sector we have two parameters, $m_{\Psi}$ and the Yukawa
coupling constant $g$. We will determine their values by fitting the nucleon mass data.

Note that in the nucleon sector there is a crucial difference with the meson sector discussed in the
previous section. The action for the bulk spinor is only 1st order. The dilaton field will not affect
the equation of motion of bulk spinors, since no integration by parts is needed here. However we will see explicitly
below that their Yukawa couplings with the bulk scalar will give us a quadratic growth soft-wall if the v.e.v
$X_0(z)\sim O(z^2)$ as required in (\ref{Xinf}).

\subsection{Nucleon spectra}
As in the above meson sectors, nucleons in this model are identified as the Kaluza-Klein modes of the 5d Dirac spinors
\begin{eqnarray}
\Psi_1(x,z)=\begin{pmatrix} \sum_n N_{1L}^{(n)}(x)f_{1L}^{(n)}(z)\,\, \\[0.2cm] \sum_n N_{1R}^{(n)}(x)f_{1R}^{(n)}(z)\,\,  \end{pmatrix}\,\,;\quad
\Psi_2(x,z)=\begin{pmatrix} \sum_n N_{2L}^{(n)}(x)f_{2L}^{(n)}(z)\,\, \\[0.2cm] \sum_n N_{2R}^{(n)}(x)f_{2R}^{(n)}(z)\,\,  \end{pmatrix}\,\,.
\end{eqnarray}
Here $N^{(n)}$ are two-component Weyl spinors, and $f^{(n)}$ are complex scalar functions.
The internal components $f^{(n)}$ should satisfy \cite{HIY}
\begin{eqnarray}
\begin{pmatrix} \p_z-\frac{\Delta^+}{z} & -v(z) \\[0.2cm] -v(z) & \p_z-\frac{\Delta^-}{z} \end{pmatrix}
\begin{pmatrix} f_{1L}^{(n)} \\[0.2cm] f_{2L}^{(n)} \end{pmatrix}=
-m_N^{(n)}\begin{pmatrix} f_{1R}^{(n)} \\[0.2cm] f_{2R}^{(n)} \end{pmatrix}\,\,, \label{EOM1}\\[0.3cm]
\begin{pmatrix} \p_z-\frac{\Delta^-}{z} & v(z) \\[0.2cm] v(z) & \p_z-\frac{\Delta^+}{z} \end{pmatrix}
\begin{pmatrix} f_{1R}^{(n)} \\[0.2cm] f_{2R}^{(n)} \end{pmatrix}=
+m_N^{(n)}\begin{pmatrix} f_{1L}^{(n)} \\[0.2cm] f_{2L}^{(n)} \end{pmatrix}\,\,. \label{EOM2}
\end{eqnarray}
Here $\Delta^{\pm}=2\pm m_{\Psi}$, and
\begin{eqnarray}
v(z)=g\,X_0(z)\,a(z)=\frac{\,g(A+Bz^2)}{\sqrt{1+C^2z^2}\,}\,\,.
\end{eqnarray}
To get the correct chiral coupling between $\Psi$'s
and baryon operators at the UV boundary, we need to impose boundary conditions \cite{HIY}
\begin{eqnarray}
f_{1L}^{(n)}(z\rightarrow0)=0\,,\quad
f_{2R}^{(n)}(z\rightarrow0)=0\,. \label{bcf0}
\end{eqnarray}
For the boundary condition at infinity, we choose
\begin{eqnarray}
f_{1R}^{(n)}(z\rightarrow\infty)=0\,,\quad
f_{2L}^{(n)}(z\rightarrow\infty)=0\,\label{bcfoo}
\end{eqnarray}
to guarantee the normalizability of eigenfunctions.
We can eliminate $(f_{1R}^{(n)},f_{2R}^{(n)})^T$ from EOM (\ref{EOM1}) and (\ref{EOM2}) as
\begin{eqnarray}
&&\begin{pmatrix} f_{1L}^{(n)''} \\[0.2cm] f_{2L}^{(n)''} \end{pmatrix}+
\begin{pmatrix} -\frac{4}{z} & 0 \\[0.2cm] 0 & -\frac{4}{z} \end{pmatrix} \begin{pmatrix} f_{1L}^{(n)'} \\[0.2cm] f_{2L}^{(n)'} \end{pmatrix} \nonumber\\[0.2cm]
&&\quad\quad +\begin{pmatrix} m_N^{(n)\,2}-\frac{m_{\Psi}^2+m_{_\Psi}-6}{z^2}-v^2 & -v' \\[0.2cm] -v' & m_N^{(n)\,2}-\frac{m_{\Psi}^2+m_{_\Psi}-6}{z^2}-v^2 \end{pmatrix}
\begin{pmatrix} f_{1L}^{(n)} \\[0.2cm] f_{2L}^{(n)} \end{pmatrix}=0\,.
\end{eqnarray}
Here $v'$ means $\p_zv$.
We can transform the above equation to a coupled Schr\"{o}dinger form. Define
$(f_{1L}^{(n)},f_{2L}^{(n)})=(z^2\chi_{1L}^{(n)},\,z^2\chi_{2L}^{(n)})$ and
$\chi^{(n)}_L=(\chi_{1L}^{(n)},\chi_{2L}^{(n)})^T$, then
\begin{eqnarray}
\chi_L^{(n)}{''}+(m_N^{(n)\,2}-V(z))\,\chi^{(n)}_L=0\,,\label{Sch}
\end{eqnarray}
where the potential matrix $V(z)$ is
\begin{eqnarray}
V(z)=\begin{pmatrix} V_{11} & V_{12} \\[0.2cm] V_{21} & V_{22} \end{pmatrix}
    =\begin{pmatrix} \frac{m_{\Psi}^2-m_{_\Psi}}{z^2}+v^2 & v' \\[0.2cm] v' & \frac{m_{\Psi}^2+m_{_\Psi}}{z^2}+v^2 \end{pmatrix}\,.\label{ptmat}
\end{eqnarray}
Note that the diagonal entries $V_{11}$ and $V_{22}$ are both $O(z^2)$ as $z\rightarrow\infty$ due to the $v^2$ term,
and the off-diagonal ones tend to a constant. Therefore the nucleons, like mesons discussed in the previous section,
also have an asymptotically linear spectrum. This is why we require $X_0(z)$ should has the behavior as in (\ref{Xinf}).
The boundary condition for $\chi^{(n)}_L$ can be deduced from that of $f^{(n)}_L$ in (\ref{bcf0}), (\ref{bcfoo})
and the EOM (\ref{EOM1}), (\ref{EOM2}) as follows
\begin{eqnarray}
\chi^{(n)}_{1L}=0\,,\quad&& \frac{m_{\Psi}}{z}\chi^{(n)}_{2L}+\chi^{(n)}_{2L}{'}=0\,,\quad   \mathrm{at}\,\, z\rightarrow 0 \,;\label{bc1}\\[0.2cm]
\chi^{(n)}_{2L}=0\,,\quad&& \frac{m_{\Psi}}{z}\chi^{(n)}_{1L}-\chi^{(n)}_{1L}{'}=0\,,\quad   \mathrm{at}\,\, z\rightarrow \infty\,.\label{bc2}
\end{eqnarray}

We use the values of $A,B,C$ or $m_q,\sigma,\gamma\,$  coming from the meson sector.
As a first step, we fix $m_\Psi$ to its classical value 5/2 which is determined by the classical dimension of
the baryon operator, and only use the parameter $g$ to fit seven spin-$\frac{1}{2}$ nucleon masses listed in PDG \cite{PDG}.
The best value is
\begin{eqnarray}
g=16.205\,.
\end{eqnarray}
The resulting mass spectra are listed in Table \ref{Nucl'}. We can see from this table that even this one-parameter fit
is not bad. Except the ground state the agreement with the data for other excited states are all within 10\%. The average error is only 6.2\%.

\begin{table}
\centering
\begin{tabular}{|c|c|c|c|c|c|c|c|}
\hline
$N$                     &  0     &   1    &  2     &  3     &  4     &  5     &   6  \\
\hline
$m_{\mathrm{exp}}$(Mev) & 940    & 1440   & 1535   & 1650   & 1710   & 2090   & 2100 \\
\hline
$m_{\mathrm{th}}$(Mev)  & 1144   & 1398   & 1509   & 1698   & 1798   & 1955   & 2046 \\
\hline
error                   &21.7\%   &2.9\%   &1.7\%   &2.9\%   &5.2\%   &6.5\%   &2.6\% \\
\hline
\end{tabular}
\caption{\small{The experimental and theoretical values of the spin-1/2 nucleon masses
in the \emph{one}-parameter fit. The average error is 6.2\%.}}\label{Nucl'}
\end{table}

Since the corresponding baryon operator may has nonzero anomalous dimension, we can treat the mass $m_\Psi$
also as a free parameter. By fitting the experimental data, the preferred values are
\begin{eqnarray}
m_\Psi=1.5\,,\qquad g=19.515\,.
\end{eqnarray}
In Table \ref{Nucl} we show the predicted values of our soft-wall nucleon model and the corresponding
experimental data. We can see that this two-parameter fit improves the agreement.
Even the worst one is smaller than 7.5\%. The average error is only 4.7\%.
\begin{table}
\centering
\begin{tabular}{|c|c|c|c|c|c|c|c|}
\hline
$N$                     &  0     &   1    &  2     &  3     &  4     &  5     &   6  \\
\hline
$m_{\mathrm{exp}}$(Mev) & 940    & 1440   & 1535   & 1650   & 1710   & 2090   & 2100 \\
\hline
$m_{\mathrm{th}}$(Mev)  & 1003   & 1333   & 1477   & 1701   & 1826   & 2004   & 2117 \\
\hline
error                   &6.7\%   &7.4\%   &3.8\%   &3.1\%   &6.8\%   &4.1\%   &0.8\% \\
\hline
\end{tabular}
\caption{\small{The experimental and theoretical values of the spin-1/2 nucleon masses in the \emph{two}-parameter fit.
The average error is 4.7\%.}}\label{Nucl}
\end{table}

\section{Summary and discussion}

In this paper we introduce a modified soft-wall AdS/QCD model in which linear confinement is realized
in both meson and nucleon sectors. Linear confinement in the meson sector require the dilaton to be
asymptotically quadratic at infinity, while in the nucleon sector it requires the v.e.v. of the bulk
scalar to be quadratic growth. The crucial point for these two conditions being satisfied at the same
time is the cubic interaction term in the bulk scalar potential.\footnote{It seems not easy to realize
this potential in a brane set-up. The quark-bilinear can be interpreted as an open string tachyon
living on the flavor brane. The potential is usually an even function of the tachyon field. The author
thanks the referee for pointing out this fact.} We also compare the predicted mass spectra with the experimental
data and find good agreement between them.

There are also some further issues, beyond the scope of this paper, need to be studied in the future. For example,
extend this model to spin-$\frac{3}{2}$ $\Delta$ resonances. More importantly, we could study the coupling
between mesons and nucleons. In the hard-wall model it seems that mesons and nucleons need different IR
cut-offs to fit the data. This problem becomes much more serious when we consider their couplings.
However in our soft-wall model presented in this paper, this consistency problem disappears by definition.
Therefore it seems more natural to study the meson-nucleon couplings in our model.

\section*{Acknowledgements}
I would like to thank Prof. Y.-C. Huang and W.-Y. Wang for many interesting discussions,
and Prof. Y.-L. Wu for his talk given in the workshop on particle physics at BJUT, which
stimulated the present work.

\end{document}